\documentclass[a4paper,twocolumn,prl,amssymb,aps, longbibliography]{revtex4-2}
\usepackage{dcolumn}
\usepackage{bm}
\usepackage{graphicx,color}
\usepackage{epstopdf}
\usepackage{xspace}
\usepackage[english]{babel}
\usepackage{verbatim}
\usepackage{braket}
\usepackage{hyperref}
\hypersetup{
    colorlinks=true,
    linkcolor=blue,
    filecolor=magenta,      
    urlcolor=cyan
    }
\usepackage{booktabs}
\usepackage{ulem} 
\usepackage[version=3]{mhchem}

\newcommand{\ac}{$\alpha_{c}$}

\newcommand{\TC}{$T_\mathrm{C}$}

\newcommand{\CGT}{Cr$_2$Ge$_2$Te$_6$}
\newcommand{\CST}{Cr$_2$Si$_2$Te$_6$}
\newcommand{\IGT}{In$_2$Ge$_2$Te$_6$}
\newcommand{\DLL}{${\Delta}L/L$}

\begin{document}
	
\title{Strong effects of uniaxial pressure and short-range correlations in Cr$_2$Ge$_2$Te$_6$}
\author{S. Spachmann$^{1,a}$, A. Elghandour$^{1}$, S. Selter$^{2}$, B. B\"uchner$^{2,3}$, S. Aswartham$^{2,b}$, R. Klingeler$^{1,c}$}

\affiliation{$^1$Kirchhoff Institute for Physics, Heidelberg University, INF 227, 69120 Heidelberg, Germany}
\affiliation{$^2$Leibniz Institute for Solid State and Materials Research (IFW), Helmholtzstr. 20, 01069 Dresden, Germany}
\affiliation{$^3$Institute of Solid State and Materials Physics and W\"{u}rzburg-Dresden Cluster of Excellence ct.qmat, Technische Universit\"{a}t Dresden, 01062 Dresden, Germany}

\begin{abstract}

\CGT\ is a quasi-two-dimensional semiconducting van der Waals ferromagnet down to the bilayer with great potential for technological applications. Engineering the critical temperature to achieve room-temperature applications is one of the critical next steps on this path. Here, we report high-resolution capacitance dilatometry studies on \CGT\ single crystals which directly prove significant magnetoelastic coupling and provide quantitative values of the large uniaxial pressure effects on long-range magnetic order (${\partial}T_{\mathrm{C}}/{\partial}p_{\mathrm{c}} = 24.7$~K/GPa and ${\partial}T_{\mathrm{C}}/{\partial}p_{\mathrm{ab}} = -15.6$~K/GPa)  derived from thermodynamic relations.
Moderate in-plane strain is thus sufficient to strongly enhance ferromagnetism in \CGT\ up to room temperature.
Moreover, unambiguous signs of short-range magnetic order up to 200~K are found.
\end{abstract}

\date{\today}

\maketitle

Strain is a versatile parameter to engineer electronic, optical, thermal, or chemical properties of materials in semiconductor technology~\cite{Chen2019, Peng2020, Chen2020, Tsutsui2019, Cenker2022}. When technologically relevant phenomena in unstrained single crystals or epitaxial films appear only at low temperatures, it can be used particularly to tune the relevant phenomena toward room temperature~\cite{Schlom2014, Wang2018ACS, Homm2021}. The quasi-two-dimensional (quasi-2D) van der Waals (vdW)  semiconducting ferromagnet \CGT\ is such an example of great potential for technological applications as it shows ferromagnetism in bilayers at around 30~K \cite{Gong2017}. The precise determination of uniaxial strain effects as demonstrated by gigantic uniaxial pressure dependencies reported here is hence mandatory to engineer \TC\ to the room temperature as a next step towards applications. 

In particular due to groundbreaking recent discoveries of long-range magnetic order in bilayers of \CGT\ \cite{Gong2017} and down to the monolayer in CrI$_3$\ \cite{Huang2017}, quasi-2D layered vdW materials are at the forefront of research. The layered structure with weak bonding between individual layers makes these materials very attractive for both fundamental research and application-oriented communities. Fundamental research on magnetic vdW materials is focused on the origin and control of their magnetic anisotropy and spin-coupling mechanisms~\cite{Zhang2016, Kim2019}, whereas research on applications ranges from integration into vdW heterostructures~\cite{Idzuchi2019, Blei2021}, vdW-materials-based spintronic devices~\cite{Zhong2017, Wang2018NatComm, Alghamdi2019}, and field effect transistors, e.g., using layered NiPS$_3$~\cite{Jenjeti2018}, to  thermoelectric devices~\cite{Tang2017}.

In bulk \CGT , ferromagnetism evolves at \TC\ $\simeq 65$~K~\cite{Zeisner2019,Selter2020}. It crystallizes in the trigonal space group $R\bar{3}$ (No.~148) and belongs to the class of layered vdW transition metal trichalcogenides (TMTCs). The edge-sharing transition metal chalcogenides form a honeycomb network. These honeycomb layers are stacked along the $c$ axis with a vdW gap between adjacent layers (Fig.~S1 of the Supplemental Material).
A key feature in the layered magnetic vdW materials is magnetic anisotropy, which enables the persistence of long-range magnetic order down to the bilayer or monolayer limit. 
It originates from the crystallographic lattice, which is connected to the electronic spins via spin-orbit coupling~\cite{Selter2020, Kim2019}. Previous studies investigated the control of magnetic anisotropy in \CGT\ by hydrostatic pressure, discussing the possibility of switching the uniaxial anisotropy under pressure from the crystallographic $c$ axis in the unstrained system to an in-plane anisotropy~\cite{Lin2018, Sakurai2020}.
Moreover, spin correlations in \CGT\ up to 160~K, i.e., about 2.5 times \TC, were inferred from static magnetic susceptibility and X-band electron spin resonance (ESR) measurements~\cite{Sun2019PCCP,Zeisner2019}, and angle-resolved photoemission spectroscopy (ARPES) spectra of \CGT\ at 150~K rather suggest a ferromagnetic than a paramagnetic state~\cite{Suzuki2019}.

These findings motivated us to investigate \CGT\ by means of high-resolution capacitance dilatometry to study lattice changes under changing temperature or magnetic field at an extremely high resolution~\cite{Kuechler2012}. What is more, in conjunction with specific heat measurements it enables the derivation of the uniaxial pressure dependencies of \TC\ from thermodynamic relations, providing the means to quantify the magnetoelastic coupling in a material. In contrast to hydrostatic pressure, uniaxial pressure dependencies are otherwise hard to access experimentally and have not been reported previously for \CGT.

Single crystals of \CGT\ have been grown by the self-flux technique and were characterized in detail as reported in Refs.~\onlinecite{Zeisner2019,Selter2020}. High-resolution dilatometry measurements were performed by means of two three-terminal high-resolution capacitance dilatometers from Kuechler Innovative Measurement Technology in a home-built setup placed inside a variable-temperature insert in an Oxford magnet system~\cite{Kuechler2012, Kuechler2017, Werner2017}. The capacitance readout was facilitated by Andeen-Hagerling's AH~2550A Ultra-Precision 1-kHz capacitance bridge~\footnote{Andeen-Hagerling Inc., AH 2550A 1 kHz Ultra-Precision Capacitance Bridge}. 
With the dilatometers, the uniaxial thermal expansion ${\Delta}L_i(T)/L_i$ and the linear thermal expansion coefficients $\alpha_i=1/L_i\times dL_i(T)/dT$ both along the $c$ axis and along the in-plane direction, i.e., $\parallel ab$, were studied at temperatures between 2 and 300~K in zero field and in magnetic fields up to 15~T applied along the direction of the measured length changes. 
In addition, the field-induced length changes ${\Delta}L_i(B_i)$ were measured at various fixed temperatures between 2 and 204~K in magnetic fields up to 15~T~\cite{Spachmann2021PhD}.
The crystals measured in this study were cut to cuboid shapes with lengths of the order of 300~$\mu$m along the $c$ axis. In-plane measurements were performed on a crystal with dimensions of $1.3\times 2.0$~mm$^2$ ($l_{ab} = 1.29$~mm) in the $ab$ plane. 
Zero-field specific heat was determined in a Physical Property Measurement System (PPMS) from Quantum Design. 


\begin{figure}[ht]
	\center{\includegraphics [width=1\columnwidth,clip]{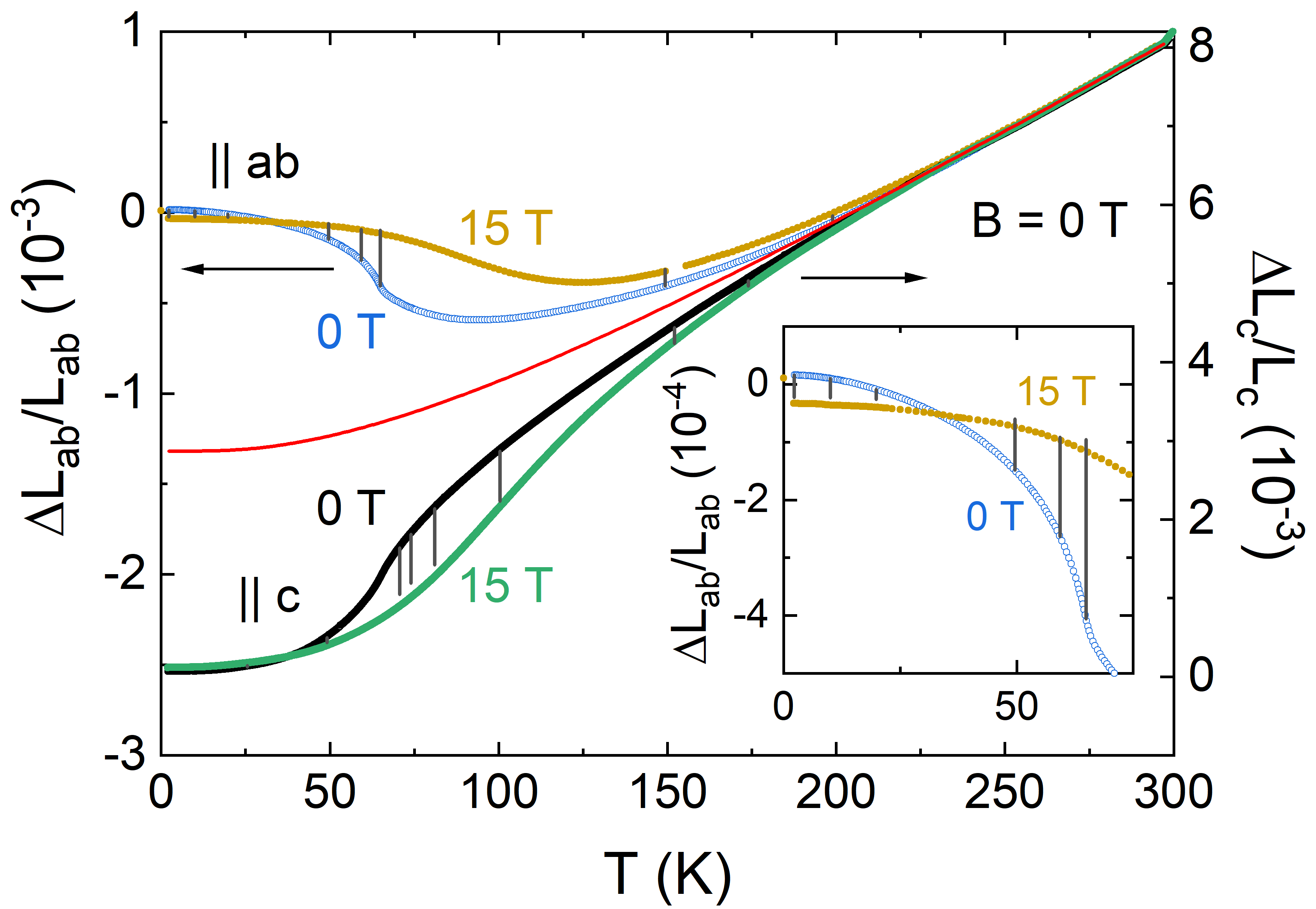}}
	\caption[] {\label{TE_0T-15T} Thermal expansion at 0 and 15~T with nonmagnetic background estimate. Relative length changes in zero field (black and blue circles) and at $B=15$~T (green and brown symbols) for $B\parallel ab$ and $B\parallel c$. Left and right ordinates have been shifted and scaled.
	The red curve shows the estimated phonon background (see the text). Vertical lines indicate the length changes between 0 and 15~T obtained from magnetostriction measurements at selected temperatures. The inset highlights the low-temperature behavior of ${\Delta}L_{ab}$($T$).
	}
\end{figure}

\begin{figure}[ht]
	\center{\includegraphics [width=1\columnwidth,clip]{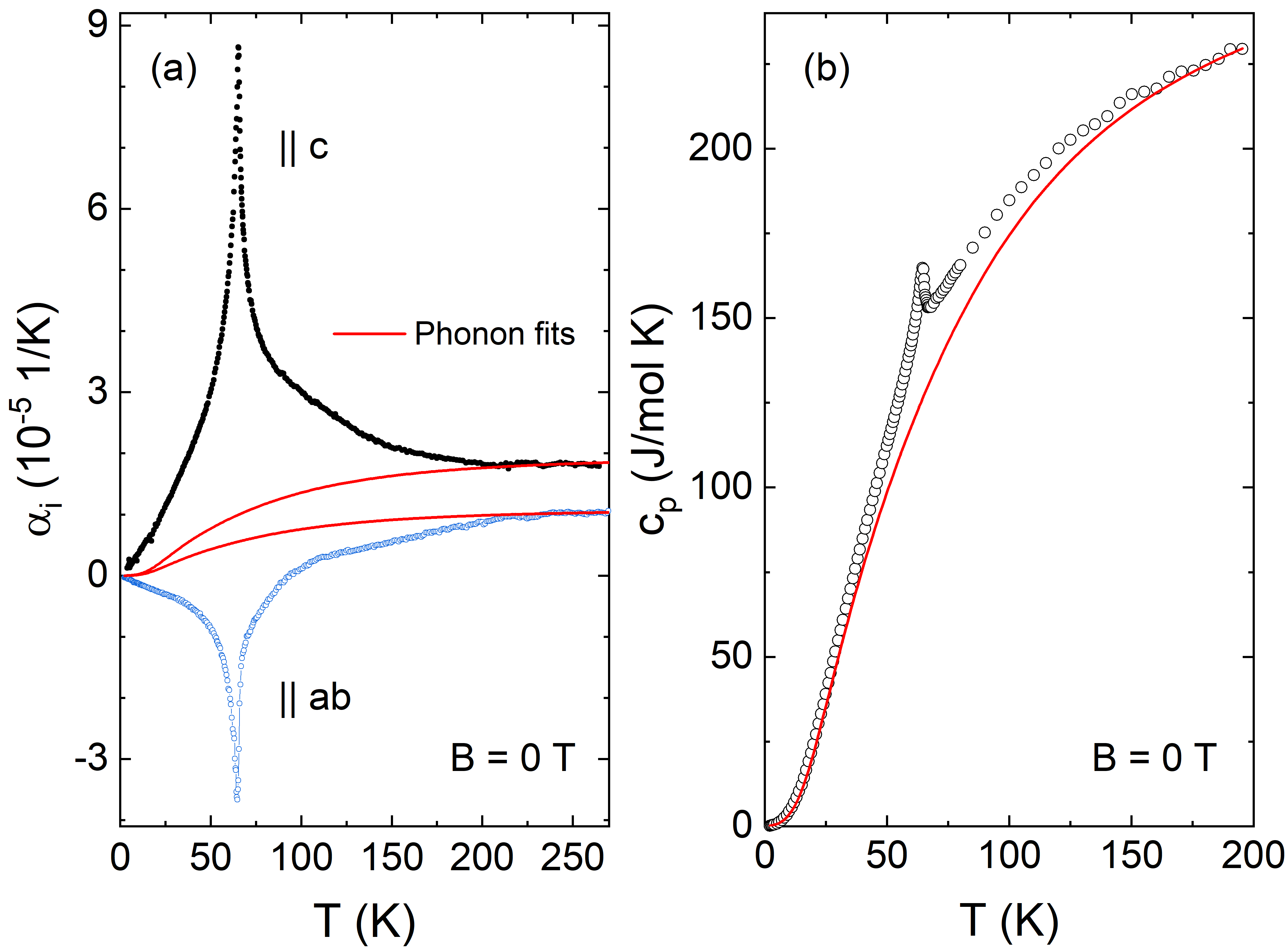}}
	\caption[] {\label{Alpha-vs-c_p} (a) Thermal expansion coefficients $\alpha_i$, $i = c, ab$ and (b) specific heat at $B=0$~T. The red curves show the background fits as explained in the text.}
\end{figure}

Thermal expansion measurements evidence strong magnetoelastic coupling in \CGT\ as shown by pronounced kinks in \DLL\ (Fig.~\ref{TE_0T-15T}) and corresponding peaks in the thermal expansion coefficients at \TC , in zero magnetic field (Fig.~\ref{Alpha-vs-c_p}(a)).
The anomalies signal a continuous phase transition with an increase (decrease) in the in-plane (out-of-plane) lattice parameters at \TC\ upon cooling. As pressure stabilizes the phase with smaller lattice, this implies negative dependence of \TC\ on pressure applied within the $ab$ plane whereas pressure applied along the $c$ axis will increase \TC. The anomalies in the thermal expansion data in Fig.~\ref{Alpha-vs-c_p} are very pronounced at \TC, but extend from the lowest measured temperatures up to at least 200~K, which can easily be seen for the $c$ axis: Phononic contributions to the thermal expansion coefficient are expected to monotonically increase with temperature, whereas \ac\ decreases until about 200~K, where it reaches a plateau. This behavior suggests a large regime of nonphononic precursory length changes above the long-range ferromagnetic ordering temperature. The precursory length changes are associated with negative in-plane thermal expansion up to nearly 95~K whereas \ac~is positive in the whole temperature regime under study~\footnote{Due to the softness of the material, in-plane sample mounting is very susceptible to pressure inevitably applied in the capacitance dilatometer and in-plane length changes display larger error bars than out-of-plane data. In particular, different mounting of the sample yields a somewhat smaller anomaly in $\alpha_{ab}$ at \TC.}

Applying an external magnetic field of 15~T yields distinct field-induced effects on the length changes (Fig.~\ref{TE_0T-15T}, green and brown symbols). Note that the 15~T data for each axis in Fig.~\ref{TE_0T-15T} are shifted vertically to coincide at highest temperatures with the respective 0~T data where magnetostriction is vanishing. The field effect at 15~T becomes distinguishable upon cooling below roughly 210~K. Experimentally determined magnetostriction ${\Delta}L(B)/L(0)$ unambiguously confirms the magnetic field effect (vertical lines in Fig.~\ref{TE_0T-15T} showing ($L$(15~T)- $L$(0))/$L$(0) at various temperatures; see also Fig.~S5). In line with this large field effect, the peak in the thermal expansion coefficients shifts to higher temperatures upon application of the magnetic field for both directions, i.e., to 90~K (in-plane) and 103~K ($c$ axis), and broadens substantially (Fig.~S3). For $B\parallel c$, the data imply shrinking of the $c$ axis at 30~K~$\lesssim T\lesssim$~210~K and only very small magnetostriction outside this temperature regime. In contrast, magnetostriction is positive for $B\parallel ab$ but changes sign at ${\sim}37$~K (see inset of Fig.~\ref{TE_0T-15T}). The magnetostriction coefficient is given by $\lambda_i = 1/L_{i}{\partial}L_{i}/\partial B = -{\partial}M/{\partial} p_{i}$, where $M$ is the magnetization and $p_i$ is the stress applied along $i$. This observation implies a sign change of the uniaxial pressure dependence of the magnetization from $\partial M/\partial p_{ab}>0$ to $\partial M/\partial p_{ab}<0$ at low temperatures. The $c$ axis pressure dependence $\partial M/\partial p_{c}$, on the other hand, is very small at lowest temperatures. 

\begin{figure}[htb]
	\center{\includegraphics [width=1\columnwidth,clip]{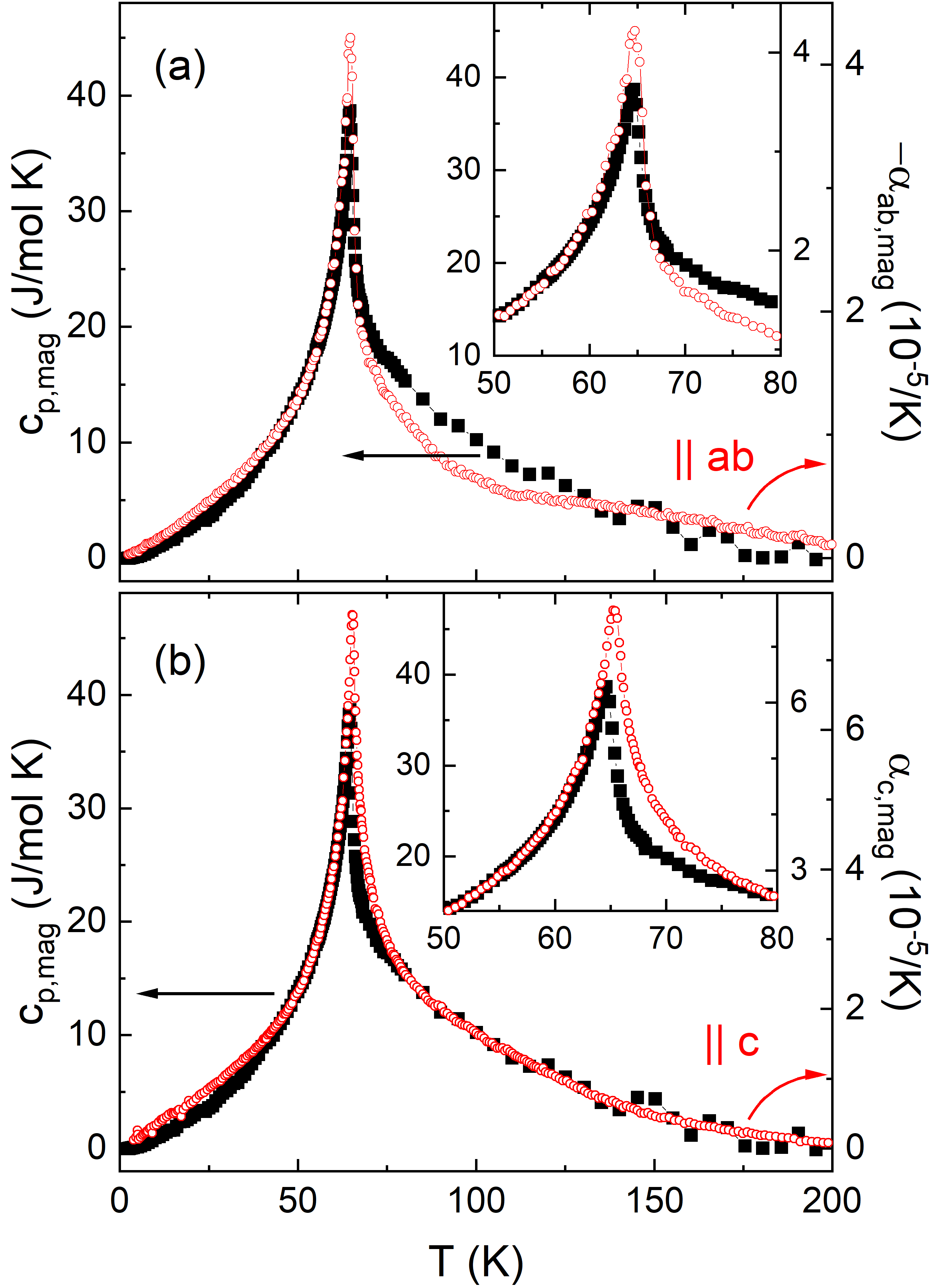}}
	\caption[] {\label{gruen2} Gr\"{u}neisen scaling of thermal expansion and specific heat. Comparison of $\alpha_{i,\mathrm{mag}}$ (red open symbols) and c$_{\mathrm{p,mag}}$ (black closed symbols) at $B=0$~T for the in-plane direction (a) and the $c$ axis (b) yielding the magnetic Gr\"{u}neisen parameters $\gamma^{\mathrm{mag}}_{\mathrm{ab}}= 4.2-7.8$ and $\gamma^{\mathrm{mag}}_{\mathrm{c}}= 11.3$ (see text).}
\end{figure}

In order to estimate nonphononic contributions to the thermal expansion of \CGT\ an estimate of the phononic contributions is necessary~\footnote{A phonon background correction using the nonmagnetic analog \IGT\ was unsuccessful, see supplemental material.}.
In general, the phononic contribution to the linear thermal expansion coefficient $\alpha_i$ is related to the specific heat of a phonon mode $j$, $c_{ph,j}$ (in J mol$^{-1}$ K$^{-1}$), by the compressibility $\kappa$ (in GPa$^{-1}$) and a Gr\"{u}neisen parameter $\gamma_{i,j}$ (dimensionless) as
\begin{equation}\label{eq:Gruen}
    \alpha_{i,ph}(T) = \frac{\kappa}{3 V_{\mathrm{m}}} \cdot \sum_{j} \gamma_{i,j} \cdot c_{ph,j}(T)
\end{equation}
where $V_{\mathrm{m}} = 1.67\times 10^{-4}$~m$^3$/mol is the molar volume for \CGT, calculated from the unit cell volume $V_0 = 811 \mathring{A}^3$~\cite{Carteaux1995} and $M_{\mathrm{mol}} = 1014.87$~g/mol with three formula units per unit cell. 
The specific heat of a phonon mode can be modeled in different ways. Often a Debye approximation (for formula, see Supplemental Material), which assumes a linear dispersion of a phonon branch, is used to model acoustic phonons and low-temperature behavior, whereas an Einstein approximation is applied when optical phonons dominate. We tried different combinations of Debye and Einstein modes. The best fit to the specific heat data (Fig.~\ref{Alpha-vs-c_p}(b)) was achieved by (1) optimizing for a total magnetic entropy in line with a spin-$\frac{3}{2}$ system, $S_{\mathrm{mag}} \approx S_{\mathrm{mag,theo}} = 2Rln(4)$ (with 2 moles of Cr atoms per mole of \CGT\ and $R$ being the molar gas constant), as well as (2) assuming that the peak shape of $c_{p,\mathrm{mag}}$ should resemble that of the thermal expansion coefficient and (3) assuming that the magnetic entropy should vanish around 200~K as indicated by the plateau in \ac\ (Fig.~\ref{Alpha-vs-c_p}(a)) as explained above.
With these assumptions we obtained the best fit for a combination of only two Debye modes with Debye temperatures $\Theta_{D,1} = 150$~K and $\Theta_{D,2} = 410$~K and a Sommerfeld coefficient $\gamma_{el} = 60$~mJ mol$^{-1}$ K$^{-2}$ (obtained from the low-temperature behavior; see~Supplemental Material).
The magnetic specific heat (Fig.~\ref{gruen2}, black squares) results from subtracting the phonon fit from $c_p$. The related magnetic entropy shows that only 56\% of the full magnetic entropy is contained below \TC.
Phonon fits to the thermal expansion coefficients (Fig.~\ref{Alpha-vs-c_p}(a)) are obtained according to Eq.~\eqref{eq:Gruen} by multiplying the phononic specific heat by a constant, the effective Gr\"{u}neisen parameter $\gamma_{i,\mathrm{eff}} = \kappa \gamma_{i}/(3 V_{\mathrm{m}})$, to fit the high-temperature behavior of $\alpha_i$. With this approach we assume a constant bulk compressibility, $\kappa(T) =$ const, as well as a constant Gr\"{u}neisen parameter which is the same for both Debye modes, $\gamma_{i,1} = \gamma_{i,2} = \gamma_i$, $i = c, ab$.
This results in $\gamma_{c,\mathrm{eff}} = 8.05(10)\times 10^{-8}$~mol/J and $\gamma_{ab,\mathrm{eff}} = 4.5(1)\times 10^{-8}$~mol/J. Bulk compressibilities for \CGT\ have not been determined experimentally, but a bulk modulus $B = 1/\kappa = 14$~GPa has been reported from density functional theory (DFT) calculations~\cite{BulkModulus}, i.e., $\kappa = 0.071$~GPa$^{-1}$. Applying this result from calculations yields an estimate of the Gr\"{u}neisen parameters $\gamma_c = 0.56$ and $\gamma_{ab} = 0.32$.
The estimated nonmagnetic background in Fig.~\ref{TE_0T-15T} is obtained by integration of these fits with respect to temperature.

As seen in Figs.~\ref{TE_0T-15T} and \ref{Alpha-vs-c_p}, above 205~K the thermal expansion is described very well by the phonon background. Upon cooling below 205~K, in contrast, nonphononic thermal expansion evolves which is anisotropic in nature. We attribute the non-phonon behavior to magnetic degrees of freedom. 
A comparison of these magnetic contributions to the specific heat and thermal expansion reveals that they scale over the full range from 0 to 200~K except at the Curie temperature (Fig.~\ref{gruen2}; see also Fig.~S4).
This scaling implies the presence of only one dominant energy scale~\cite{Gegenwart2016Grueneisen, Klingeler2006Pressure}.
While the $c$ axis shows a nearly perfect scaling from 200 to 80~K and good scaling below 64~K, minor deviations can be seen for the $ab$ plane, especially between 70 and 130~K.
The effective magnetic Gr\"{u}neisen parameter for the $c$ axis is $\gamma_{c,\mathrm{mag,eff}} = 1.6\times 10^{-6}$~mol/J and $\gamma_{ab,\mathrm{mag,eff}}$ varies from $-6\times 10^{-7}$~mol/J to $-1.1\times 10^{-6}$~mol/J between 120 and 30~K. With $\kappa = 0.071$/GPa as before this yields large magnetic Gr\"{u}neisen factors $\gamma_{c,\mathrm{mag}} = 11.3$ and $\gamma_{ab,\mathrm{mag}} = 4.2-7.8$.
These values are of the same magnitude as other magnetic Gr\"{u}neisen factors, e.g., $\gamma_{\mathrm{mag}} \approx -18$ for $\alpha$-Mn~\cite{White2015}.
A part of the deviations around \TC\ stems from the fact that the specific heat was measured by a relaxation method and thus averages over a temperature range, whereas the thermal expansion data are measured closer to thermal equilibrium, with a warming rate of 0.3~K/min. Effects of uniaxial pressure and strain, as well as domain formation, presumably contribute further to the deviations. Furthermore, the bulk modulus may change around \TC\ and lead to additional differences between thermal expansion and specific heat.

\begin{figure}[htbp]
	\center{\includegraphics [width=1\columnwidth,clip]{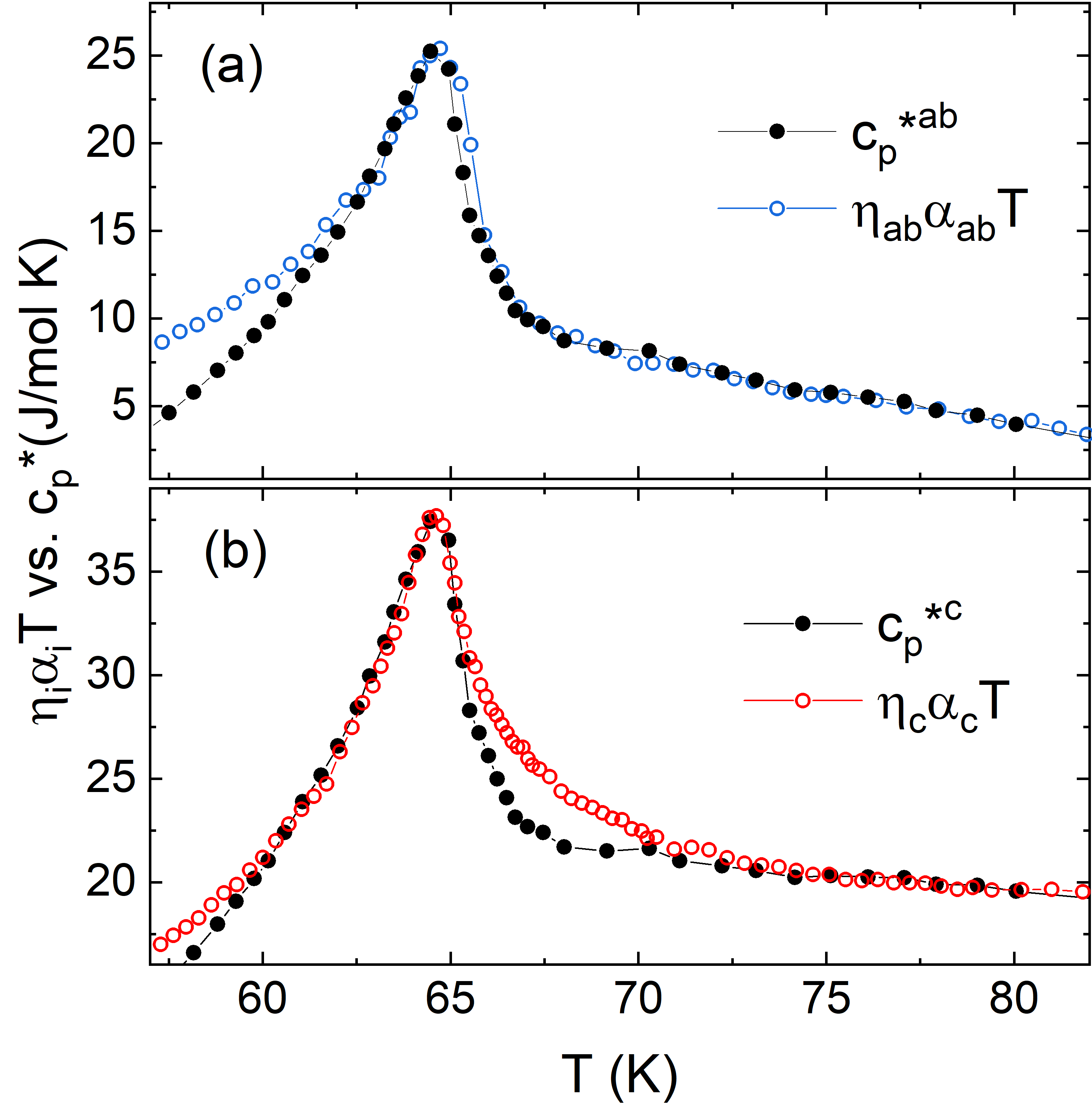}}
	\caption[] {\label{Souza} Scaling of the specific heat and thermal expansion coefficients in order to obtain the uniaxial pressure dependence according to Ref.~\onlinecite{Souza2005}. The thermal expansion coefficients were rescaled by (a) $\eta_{ab} = -10700$~J mol$^{-1}$ K$^{-1}$ and (b) $\eta_c = 6750$~J mol$^{-1}$ K$^{-1}$. $\eta_{c}\alpha_{c}$T was shifted by $-0.8$~K so that its peak position overlaps with the peak position of $c_p^{*c}$. Background subtractions with $a = 48$~J mol$^{-1}$ K$^{-1}$, $b = 1.42$~J mol$^{-1}$ K$^{-2}$ ($a = 50$~J mol$^{-1}$ K$^{-1}$, $b = 1.2$~J mol$^{-1}$ K$^{-2}$) were performed on $c_p$ to obtain $c_p^{*ab}$ ($c_p^{*c}$).}
\end{figure}

On top of the analysis of Gr\"{u}neisen scaling, the uniaxial pressure dependence of the ordering temperature can be determined by the method suggested by Souza $et al.$~for continuous phase transitions exhibiting critical behavior~\cite{Souza2005}. This method has been applied to a number of materials, and was verified in comparison to experimental pressure dependencies by studies on LaMnO$_3$ and CaMnO$_3$~\cite{Zhou2003,Souza2010}, as well as Na$_{x}$CoO$_2$~\cite{Sushko2005,Wooldridge2006,DosSantos2006}.
The pressure dependence is obtained by subtracting from the specific heat a constant offset as well as a linear term to obtain
\begin{equation}
    c_p^* \equiv c_p - a - bT
\end{equation}
and subsequently superimposing $c_p^*$ with the rescaled thermal expansion coefficients $\eta_i \alpha_i T$ (in J mol$^{-1}$ K$^{-1}$) in a small temperature regime around \TC~
\footnote{Note that the temperature for $\alpha_{c}$ was rescaled by subtracting 0.8~K, because the peak value is reached at 64.7~K for the specific heat data and $\alpha_{ab}$ whereas it is reached at 65.5~K for \ac.}.
The uniaxial pressure dependence is then given by
\begin{equation}
    \frac{dT_C}{dp_i} \equiv \left(\frac{dp_i}{dT}\right)_C^{-1} = \frac{\eta_i \alpha_i T}{c_p^*} \quad .
\end{equation}
Analyzing our data this way we obtain scaling factors $\eta_c = 6750(400)$~J mol$^{-1}$ K$^{-1}$ and $\eta_{ab} = -10700(600)$~J mol$^{-1}$ K$^{-1}$ (Fig.~\ref{Souza}). From ${\partial}T_{\mathrm{C}}/{\partial}p_{i} = \frac{V_{\mathrm{m}}}{\eta_{i}}$ we then obtain ${\partial}T_{\mathrm{C}}/{\partial}p_{c} = 24.7(1.8)$~K/GPa and ${\partial}T_{\mathrm{C}}/{\partial}p_{ab} = -15.6(1.1)$~K/GPa. Assuming that the hydrostatic pressure dependence can be calculated by superimposing the uniaxial ones in the three main directions yields an estimate of $dT_{\mathrm{C}}/dp_{p{\rightarrow} 0} = 2\cdot {\partial}T_{\mathrm{C}}/{\partial}p_{ab} + {\partial}T_{\mathrm{C}}/{\partial}p_{c} = -6.5(8)$~K/GPa. 


Our observation of nonphononic lattice changes up to 200~K visible in the thermal expansion (Fig.~\ref{Alpha-vs-c_p}) confirms the presence of short-range magnetic order far above \TC\ and provides a macroscopic measurement of high temperature magneto-structural correlations in \CGT. This observation strongly supports the expected quasi-2D nature of magnetism in \CGT\ and suggests low-dimensional exchange interactions~\cite{Calder2021,Williams2015}.
While the lattice response has not been resolved before, deviations from purely paramagnetic behavior up to around 150~K were previously observed in 
the static susceptibility~\cite{Sun2019PCCP}, in X-band ESR spectra~\cite{Zeisner2019}, and in ARPES~\cite{Suzuki2019}. Moreover, temperature dependent Raman spectra of \CGT\ show a mode around 122~cm$^{-1}$ broadening strongly upon cooling between 200 and 250~K~\cite{Sun2018ApplPhysLett}. Direct investigation of magnetic correlations on the isostructural ferromagnet \CST, which also possesses uniaxial $c$-axis magnetic anisotropy, revealed static and dynamic magnetic in-plane correlations at least up to 300~K, almost ten times the long-range ordering temperature \TC\ $= 33$~K~\cite{Williams2015}.


Low-dimensional high-temperature correlations are especially of interest from a fundamental perspective. The effects of pressure on the critical temperatures and other physical properties in magnetic vdW compounds, on the other hand, are a field of intense research with an eye on precisely tuning material properties for technological applications. Accordingly, the effects of stress and strain on \CGT~\cite{Sun2018ApplPhysLett, Sakurai2020, Fumega2020} and related materials~\cite{Chen2015, Zhao2018, Huang2019, Leon2020, Wang2020, Xue2020, Zhang2021, Zhu2021} have been intensively studied both experimentally and numerically. Previous experimental studies on \CGT\ were restricted to hydrostatic pressure $p_h$, the effect of which on the layered structure is not \textit{a priori} clear and is supposed to mainly modify the interlayer structure~\cite{Li2019}. Hydrostatic pressure is found to decrease \TC , the response being stronger for small applied pressure than for larger pressures. In the ranges from 0 to 0.25~GPa~\cite{Sun2018ApplPhysLett} and from 0 to 0.1~GPa~\cite{Fumega2020} a drop in \TC\ of about $-14$~K/GPa was observed, compared with about $-4$~K/GPa for pressures of 1~GPa and above~\cite{Sakurai2020}, which is in good agreement with $dT_{\mathrm{C}}/dp_{p{\rightarrow} 0} = -6.5(8)$~K/GPa derived from our thermal expansion and specific heat data. A more recent study found smaller changes of about $-12$~K between 0 and 4.5~GPa, i.e., a drop by about $2.7$~K/GPa~\cite{Bhoi2021}.
The changes in \TC\ under hydrostatic pressure have been ascribed to the interplay of a decrease in Cr--Cr and Cr--Te bond lengths and an increase in Cr--Te--Cr bond angle away from 90$^{\circ}$ upon increasing $p_h$, i.e., the complex competition between antiferromagnetic (AFM) direct exchange between Cr ions and superexchange mediated by the Te ions~\cite{Goodenough1963, Chen2015, Sakurai2020}. In this respect, our observation that (uniaxial) in-plane pressure yields a drastic decrease in \TC\ is particularly relevant as it suggests that the observed $dT_{\mathrm{C}}/dp_h$ is much more moderate due to the out-of-plane pressure components. This conclusion is supported by recent numerical studies of in-plane and out-of-plane exchange couplings, $J_{\mathrm{in}}$ and $J_{\mathrm{out}}$, which show a strong increase in $J_{\mathrm{out}}$ upon application of hydrostatic pressure, with an initial effect of about 35\%/GPa, whereas $J_{\mathrm{in}}$ was found to decrease by about $-10$\%/GPa~\cite{Fumega2020}. The data presented here are also consistent with recent density functional theory calculations on \CST\ where a strong enhancement of \TC\ by in-plane biaxial tensile strain was found~\cite{Chen2015}.

Our study {\it experimentally} evidences and {\it quantifies} by thermodynamic relations large uniaxial pressure effects in \CGT. These large effects underline the feasibility of using the material for potential applications, as tailoring of the transition temperature is made possible by applying moderate in-plane tensile strain. Especially, enhancing \TC\ with the goal of room-temperature applications is essential for actual devices, e.g., for sensing, data storage, or computing. The practical feasibility of such \TC\ enhancement in \CGT\ has been shown by Wang $et al.$, who managed to enhance \TC\ to 208~K by electrochemical intercalation of organic molecules into the van der Waals gap~\cite{Wang2019}. 
Adding to this result the large in-plane pressure dependence which we report here, in particular, suggests an exceptional feasibility of using \CGT\ for strain-tailoring ferromagnetism.

Although there are quite a number of theoretical studies on the effects of strain and stress on \CGT\ and other magnetic vdW materials as mentioned above, experimental results for the uniaxial pressure dependencies of these materials are very rare. 
To the best of our knowledge such measurements only exist for CrI$_3$~\cite{Arneth2022} and $\alpha$-RuCl$_3$~\cite{He2018}.
CrI$_3$ also exhibits ferromagnetic layers but shows a much smaller uniaxial pressure dependence of only ${\partial}T_{\mathrm{C}}/{\partial}p_{ab} = -0.4(1)$~K/GPa~\cite{Arneth2022}, i.e., of the same sign, but drastically, by more than one order of magnitude, smaller than in \CGT.
In contrast, the Kitaev spin liquid candidate $\alpha$-RuCl$_3$ exhibits uniaxial pressure dependencies for its different transitions of ${\partial}T_{\mathrm{N}}/{\partial}p_{a} = -3.5$~K/GPa to ${\partial}T_{\mathrm{N}}/{\partial}p_{c} = -14.5$~K/GPa. These values are much larger than those found for CrI$_3$ and yet still about two to four times smaller than what we report here for \CGT.
Note that this drastic difference is not reflected in the initial hydrostatic pressure dependencies of either CrI$_3$ or $\alpha$-RuCl$_3$ which are reported to be of similar size to that in \CGT, $dT_{\mathrm{C}}/dp = 12$~K/GPa~\cite{Mondal2019} and $dT_{\mathrm{N}}/dp \approx -13$ to $-24$~K/GPa~\cite{He2018}, respectively. 
In order to further emphasize the size of pressure dependencies found in \CGT, it is instructive to compare our results also with hydrostatic pressure dependencies of other quasi-2D magnetic materials with a honeycomb lattice, such as VI$_3$~\cite{Valenta2021}, FePS$_3$~\cite{Coak2021}, Na$_3$Ni$_2$SbO$_6$~\cite{Werner2017}, and $\beta$-Li$_2$IrO$_3$~\cite{Majumder2018}. These materials exhibit a large range of pressure dependencies differing by two orders of magnitude. Very small hydrostatic pressure dependencies have been reported for the vdW compound VI$_3$~\cite{Son2019, Valenta2021}, with $dT_{\mathrm{C}}/dp \approx 0$ at small pressures, and for Na$_3$Ni$_2$SbO$_6$, with $dT_{\mathrm{N}}/dp = -0.05$~K/GPa~\cite{Werner2017}.
Dilatometric studies of the Kitaev hyperhoneycomb iridate $\beta$-Li$_2$IrO$_3$ yield larger values of $dT_{\mathrm{N}}/dp = 0.7$~K/GPa~\cite{Majumder2018}.
Finally, a hydrostatic pressure dependence of $dT_{\mathrm{N}}/dp = 7.7$~K/GPa, is observed in the vdW antiferromagnet FePS$_3$~\cite{Coak2021}.

In conclusion, the quasi-2D layered van der Waals compound \CGT\ shows strong magnetoelastic coupling giving rise to large uniaxial pressure dependencies of \TC . These results confirm that moderate in-plane tensile strain is sufficient to strongly enhance the long-range ferromagnetic ordering temperature. Our high-resolution thermal expansion data in addition unambiguously prove short-range magnetic order up to 200~K.
The large uniaxial pressure effects and quasi-2D nature of magnetism in \CGT\ present an intriguing playground for \CGT-based technological applications, bringing into reach room-temperature ferromagnetism in 2D materials. 


We acknowledge financial support by BMBF via the project SpinFun (13XP5088) and by Deutsche Forschungsgemeinschaft (DFG) under Germany’s Excellence Strategy EXC2181/1-390900948 (the Heidelberg STRUCTURES Excellence Cluster) and through Projects No. KL 1824/13-1 (R.K.) and No. AS 523/4-1 (S.A.). B.B. acknowledges the W\"{u}rzburg-Dresden Cluster of Excellence on Complexity and Topology in Quantum Matter – ct.qmat (EXC 2147, Project No. 390858490).


\bibliography{Cr2Ge2Te6_Paper_bibliography_PRL}

\end{document}


\title{Supplemental Material: Strong effects of uniaxial pressure and short-range correlations in Cr$_2$Ge$_2$Te$_6$}
\author{S. Spachmann$^{1,}$\footnote{sven.spachmann@kip.uni-heidelberg.de}, A. Elghandour$^{1}$, S. Selter$^{2}$, B. B\"uchner$^{2,3}$, S. Aswartham$^{2,}$\footnote{s.aswartham@ifw-dresden.de}, R. Klingeler$^{1,}$\footnote{klingeler@kip.uni-heidelberg.de}}

\affiliation{$^1$Kirchhoff Institute for Physics, Heidelberg University, INF 227, 69120 Heidelberg, Germany}
\affiliation{$^2$Leibniz Institute for Solid State and Materials Research (IFW), Helmholtzstr. 20, 01069 Dresden, Germany}
\affiliation{$^3$Institute of Solid State and Materials Physics and W\"{u}rzburg-Dresden Cluster of Excellence ct.qmat, Technische Universit\"{a}t Dresden, 01062 Dresden, Germany}

\date{\today}

\maketitle

\noindent In the following
\begin{itemize}
    \item the crystal structure
    \item specific heat data and its background subtraction for \CGT\ and the non-magnetic analog \IGT\
    \item the thermal expansion coefficients at 0 and 15~T
    \item effective magnetic Grüneisen parameters
    \item high temperature magnetostriction data
\end{itemize}
are shown supporting the results of our manuscript.

\section{Crystal Structure}
The crystal structure of \CGT\ is shown in Fig.~\ref{SI_Lattice}.
\renewcommand{\thefigure}{S1}
\begin{figure*}[htbp]
	\center{\includegraphics [width=0.9\columnwidth,clip]{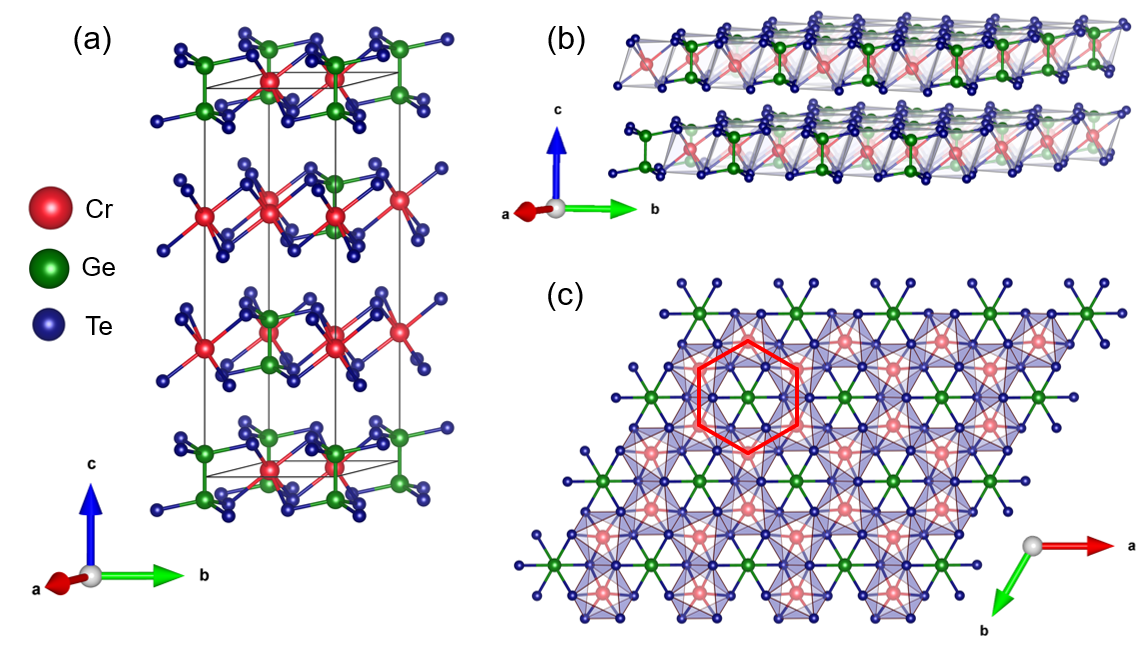}}
	\caption[] {\label{SI_Lattice} Crystal structure of \CGT\ in the space group $R\bar{3}$ (no.~148) as reported in Ref.~\cite{Carteaux1995}. (a) Unit cell of \CGT. (b) Van der Waals layers and stacking along the $c$ axis with Cr octahedra shown. (c) View onto the $ab$ plane with one unit of the honeycomb network indicated by the red hexagon.}
\end{figure*}

\clearpage
\section{Specific Heat Data}
As explained in the main text, the thermal expansion data (Fig.~2) 
suggest -- without any assumption about phonon-backgrounds or the like -- that short-range correlations in \CGT\ extend to at least 200~K.
This makes a reliable background correction difficult since the extent of the anomaly in \TC\ can not be determined easily.
Any phononic background fit to the specific heat or thermal expansion coefficient of \CGT\ will thus contain a large uncertainty.
Trying to improve the uncertainty of a background fit, we measured the specific heat of \IGT, a non-magnetic analog to \CGT (Fig.~\ref{SI_cp}(a)). The background fitting procedure to \IGT\ and \CGT\ is described in the following.
\renewcommand{\thefigure}{S2}
\begin{figure*}[htbp]
	\center{\includegraphics [width= 1\columnwidth,clip]{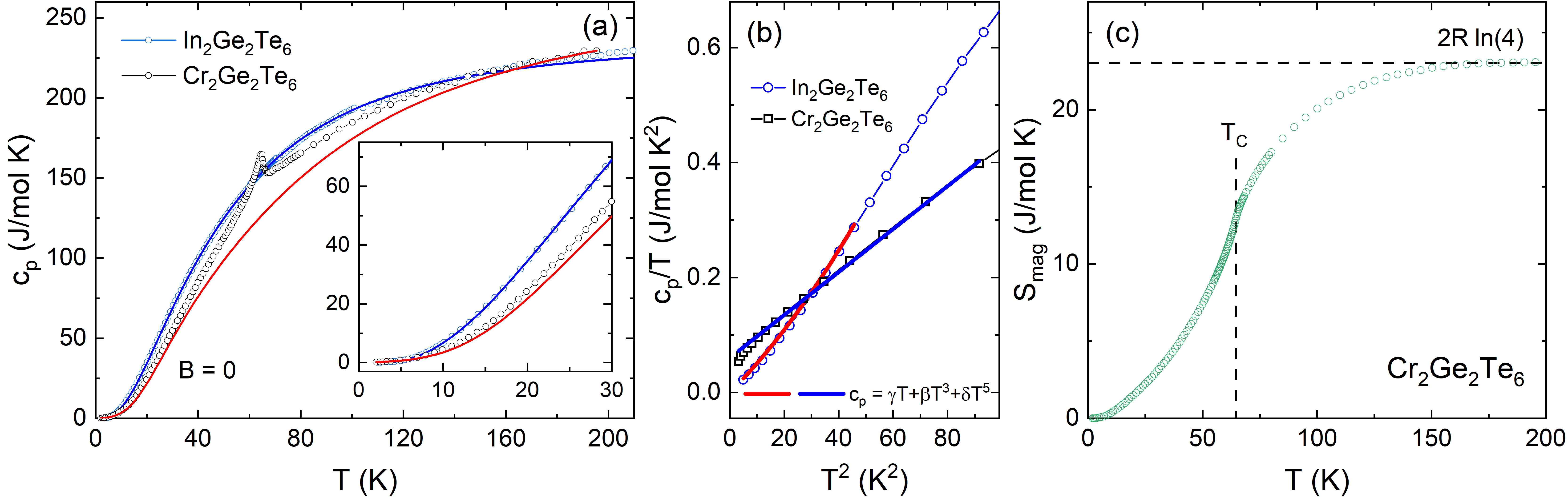}}
	\caption[] {\label{SI_cp} (a) Specific heat $c_{\mathrm{p}}$ of \IGT\ (blue open circles) and \CGT\ (black open circles) with a fits of Debye and Einstein modes (solid lines) as explained in the text. Inset: magnification of the low temperature regime. (b) Specific heat $c_{\mathrm{p}}$/T vs. T$^2$ of \IGT\ (blue open circles) and \CGT\ (black open squares) at low temperatures. Lines in (b) indicate fits to obtain $\gamma_{\mathrm{el}}$ and $\beta$ (see text). (c) Magnetic entropy of \CGT\ calculated after the phononic background subtraction from $c_{\mathrm{p}}$. The horizontal dashed line indicates the theoretically expected magnetic entropy for a spin-$\frac{3}{2}$ system. \TC\ is indicated by the vertical dashed line.}
\end{figure*}

In principle, separating out the phononic and electronic contributions from the specific heat of such an analog should enable determining the excess specific heat, i.e., the magnetic contributions, of the magnetic compound.
The phonon part of the specific heat can usually be modeled well with a combination of Debye and Einstein modes:
\begin{equation}\label{eq:DebyeEinstein}
    c_{V,ph}= \sum_i c_{V,ph,i}^{D} + \sum_j c_{V,ph,j}^{E} 
\end{equation}
with the Debye part given as    
\begin{equation}\label{eq:Debye}
    c_{V,ph,i}^{D}= 9 \cdot n_{D,i} \cdot k_B \left(\frac{T}{\Theta_{D,i}}\right)^3 \cdot \int_0^{\Theta_{D,i}/T}{\frac{x^4 e^{x}}{(e^x-1)^2} dx}
\end{equation}
and the Einstein part given by
\begin{equation}\label{eq:Einstein}
    c_{V,ph,j}^{E} = 3 \cdot n_{E,j} \cdot k_B \left(\frac{\Theta_{E,j}}{T}\right)^2 \cdot \frac{e^{\Theta_{E,j}/T}}{[e^{\Theta_{E,j}/T}-1]^2} \quad .
\end{equation}
At sufficiently low temperatures the measured specific heat at constant pressure, $c_p$, is approximately the same as the specific heat at constant volume, $c_V$,
\begin{equation}
    c_p \approx c_V \quad .
\end{equation}
The difference between c$_{\mathrm{p}}$ and c$_{\mathrm{V}}$ is given by,
\begin{equation}
    c_{p} - c_{V} = TVB\beta^2
\end{equation}
with a molar volume $V = V_{\mathrm{m}} = 1.67\times 10^{-4}$ m$^3$/mol, the volume expansion $\beta \propto 10^{-5}/$K and the bulk modulus $B \approx 14$~GPa (\CGT, see Ref.~\onlinecite{BulkModulus}), such that
\begin{equation}
    c_{p} - c_{V} < 0.1~ \mathrm{J/(mol~K)~@300~K}.
\end{equation}
Deviations between $c_p$ and $c_V$ are thus roughly on the order of 0.1~J/(mol K) around room temperature and even smaller as the temperature is decreased.

\subsection{Low Temperature Fits}
For an estimate of (1) the (linear) electronic contribution to the specific heat as well as (2) the Debye temperature we performed a low-temperature fit (Fig.~\ref{SI_cp}(b)). For this fit we used the formula
\begin{equation}
    c_p = \gamma T + \beta T^3 + \delta T^5
\end{equation}
where $\gamma = \gamma_{el}$ is the Sommerfeld coefficient and $\beta$ and $\delta$ are low temperature lattice contributions to the specific heat. $\beta$ can be used to calculate an estimate of the Debye temperature by 
\begin{equation}
    \Theta_D = \left(\frac{12 \pi^4 n R}{5 \beta} \right)^{1/3}
\end{equation}
with $R$ being the molar gas constant and $n$ the number of atoms per formula unit.

Fitting the specific heat of \IGT\ in the temperature range from 2~K to 7~K (Fig.~\ref{SI_cp}(b)) we obtained $\gamma = 0$, $\beta = 4.72$~mJ/(mol K$^4$) and $\delta = 36.6$~$\mu$J/(mol K$^{6}$). From $\beta$ we then obtained $\Theta_D = 160.3$~K.
A fit in the range from 2~K to 10~K yielded $\gamma = 0$, $\beta = 5.74$~mJ/(mol K$^4$) ($\Theta_D = 150.2$~K) and $\delta = 11.7$~$\mu$J/(mol K$^{6}$), but with a worse agreement to the data at low temperatures.

A fit to the \CGT\ specific heat from 1.8~K to 10.5~K yielded $\gamma = 62(5)$~mJ/(mol K$^2$), $\beta = 3.66$~mJ/(mol K$^4$) ($\Theta_D = 174.5$~K) and $\delta = 0$. For comparison, at fit from 1.8~K to 14.5~K yielded $\gamma = 91(8)$~mJ/(mol K$^2$), $\beta = 3.23$~mJ/(mol K$^4$) ($\Theta_D = 181.9$~K) and also $\delta = 0$, but with a worse fit quality.

From these low-temperature fits we adopted the Sommerfeld coefficients $\gamma_{el} = 0$ for \IGT\ and $\gamma_{el} = 60$~mJ/(mol K$^2$) for \CGT\ for the fitting up to high temperatures which is shown in the next sections.

\subsection{\IGT\ High Temperature Fitting}
The \IGT\ specific heat data in Fig.~\ref{SI_cp}(a) shows a minor step around 102~K, indicating a small negative offset of all data points above 102~K, probably from a slight decoupling of the sample from the calorimeter. Therefore, the data was only fitted in the range from 2~K to 100~K. The Best fit was achieved using two Debye modes (Eq.~\eqref{eq:Debye}) and one Einstein mode (Eq.~\eqref{eq:Einstein}) with $\Theta_{D,1} = 134.4$~K, $n_{D,1} = 4.58$, $\Theta_{D,2} = 286.5$~K, $n_{D,2} = 4.47$, $\Theta_{E} = 44.5$~K and $n_E = 0.459$. The sum over the weights $n_i$ is 9.51, close to the expected value of 10 for 10 atoms per formula unit. As seen in Fig.~\ref{SI_cp}(a) and (b) this fit describes both the low-temperature and the high temperature ranges very well.

\subsection{\CGT\ High Temperature Fitting}
Bouvier et al.~\cite{Bouvier1991} suggested a simple scaling of the specific heat data of two ternary compounds by the ratio of their Debye temperatures.
However, this scaling is only valid for low temperatures, up to roughly $T \lesssim\Theta_{\mathrm{D}}/10$. Accordingly, the specific heat of \CGT\ crosses the specific heat of \IGT\ around 120~K and it is obvious that such a simple scaling by $\Theta_{\mathrm{D,IGT}}/\Theta_{\mathrm{D,CGT}} = 0.952$ fails. 
The intended \CGT\ background subtraction by the non-magnetic analog \IGT\ is thus not possible.
Instead, we tried several combinations of Debye and Einstein modes before we ended up with the combination of two Debye modes with fixed $\gamma = 60$~mJ/(mol K$^2$) as described in the main text.
For fits with the different combinations of Debye and Einstein modes, the resulting Debye and Einstein temperatures strongly depend on the applied fitting range and vary roughly around $60$~K~$< \Theta_{D,1} < 200$~K and $250$~K~$< \Theta_{D,2} < 400$~K.
The Debye temperatures of \IGT\ together with the three assumptions mentioned in the main text served as a starting point to arrive at the final values. For these final values the Debye temperatures and weights $n_{D,i}$ were incrementally changed to achieve (1) a total magnetic entropy in line with a spin-$\frac{3}{2}$ system, i.e., $S_{\mathrm{mag}} \approx S_{\mathrm{mag,theo}} = 2$R$ln(4) = 23.05$~J/(mol K) (with two moles of Cr atoms per mole of \CGT, and R being the molar gas constant), (2) a peak shape of $c_{p,\mathrm{mag}}$ resembling that of the thermal expansion coefficient and (3) a magnetic entropy vanishing around 200~K as indicated by the plateau in \ac\ (Fig.~2(a)).
The final best fit values we arrived at were $\Theta_{D,1} = 150$~K, $n_{D,1} = 4.8$, $\Theta_{D,2} = 410$~K, and $n_{D,2} = 5.024$.

The magnetic entropy resulting after subtraction of the phononic and electronic fit is shown in Fig.~\ref{SI_cp}(c).

Note that both the \IGT\ and the \CGT\ data do not reach the expected classical Dulong-Petit limit of $c_{p} = 3nR \approx 249.4$~J/(mol K) at high temperatures, where $n$ is the number of atoms per formula unit and $R$ is the molar gas constant. The \IGT\ data reach 233~J/(mol K) at 230~K and the \CGT\ data reach about 230~J/(mol K) at 195~K, with both curves still increasing. Looking at $\sum_i n_i$ from the fit results, the experimental error can thus be estimated to amount to about 5\% for \IGT\ and about 2\% for \CGT.


\section{Thermal Expansion Coefficients at 0 and 15~T}
The thermal expansion coefficients derived from the relative length changes in Fig.~1 in the main text are presented in Fig.~\ref{Alpha}. The effective Gr\"{u}neisen parameter $\gamma_{c,\mathrm{eff}} = 7.0\times 10^{-8}$ is slightly different from the one extracted from the mini-dilatometer in Fig.~2(a) ($8.05\times 10^{-8}$).
\renewcommand{\thefigure}{S3}
\begin{figure*}[htb]
	\center{\includegraphics [width=0.6\columnwidth,clip]{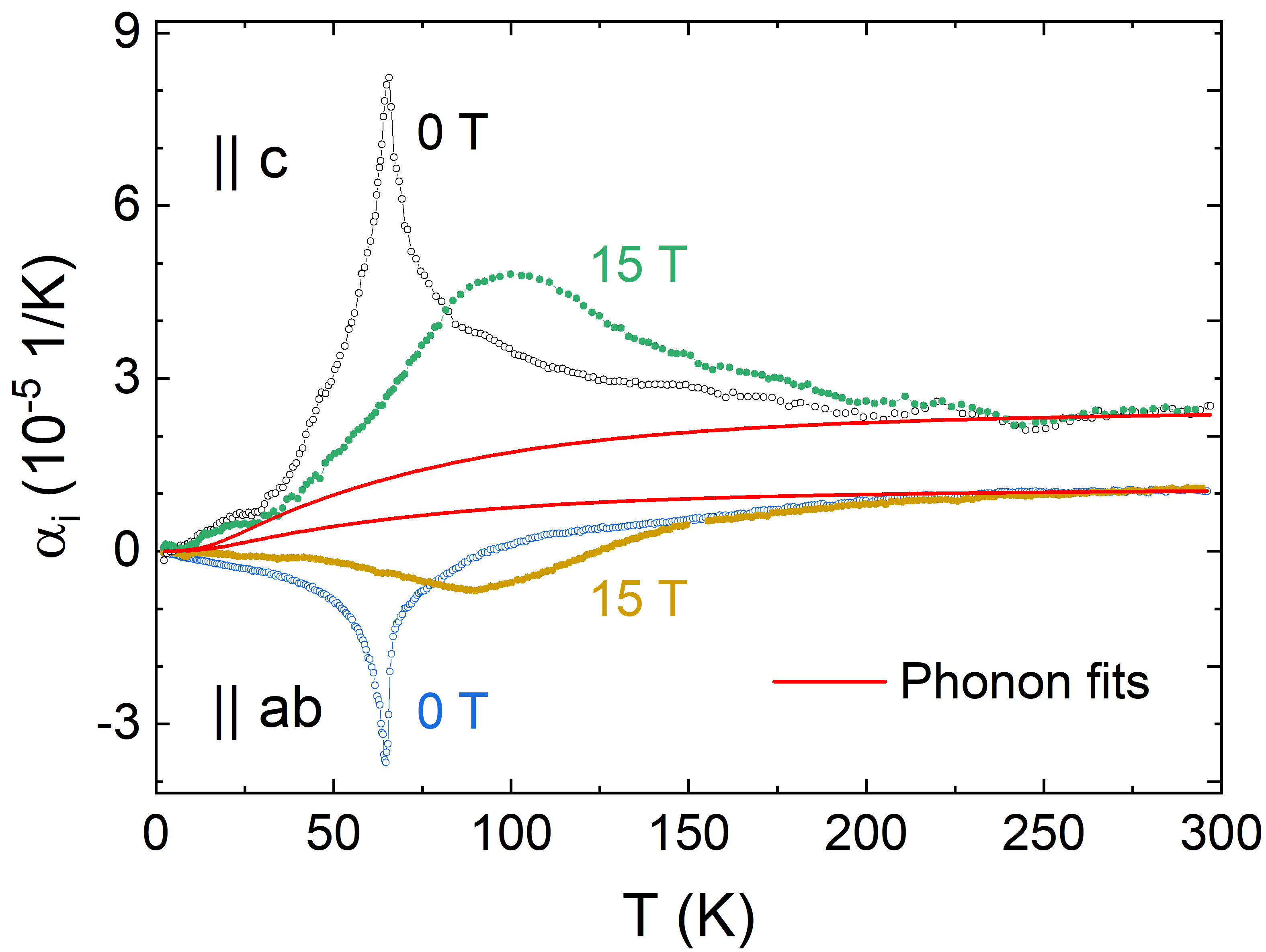}}
	\caption[] {\label{Alpha} Thermal expansion coefficients $\alpha_{i}$ of \CGT\ at 0~T and 15~T derived from the data in Fig.~1, measured in the standard dilatometer.~\cite{Kuechler2012}}
\end{figure*}

\clearpage
\section{Effective Magnetic Gr\"{u}neisen factors}
The absolute value of the effective magnetic Gr\"{u}neisen factors, $\gamma_{i,\mathrm{mag, eff}} = |\alpha_{i,\mathrm{mag}}|/c_{p,\mathrm{mag}} = \kappa \gamma_{i,\mathrm{mag}}/(3V_m)$ up to 130~K is shown in Fig.~\ref{Gruen_mag}. Above 130~K the absolute values of $\alpha_{i,\mathrm{mag}}$ and $c_{p,\mathrm{mag}}$ become small which leads to large error bars and fluctuations in the data, which is why they are not shown. 
Along the $c$ axis (red closed circles) a nearly constant value is assumed between 40~K and 130~K except between 64~K and 76~K, i.e., at and just above \TC. In this latter temperature regime a sharp peak can be seen with a jump on the low temperature side and a tail on the high temperature side. Below 40~K $\gamma_{c,\mathrm{mag, eff}}$ rises strongly.

$\gamma_{ab,\mathrm{mag, eff}}$ (black closed circles) exhibits a nearly constant value only between 90~K and 125~K, below which it has a rising "background". Here also a rise can be seen in $|\gamma_{ab,\mathrm{mag, eff}}|$ around \TC, but of opposite behavior. The tail is on the low temperature side (roughly 53~K to 65.3~K) whereas the jump can be seen on the high temperature side, between 65.3~K and 67~K.

\renewcommand{\thefigure}{S4}
\begin{figure*}[ht]
	\center{\includegraphics [width=0.6\columnwidth,clip]{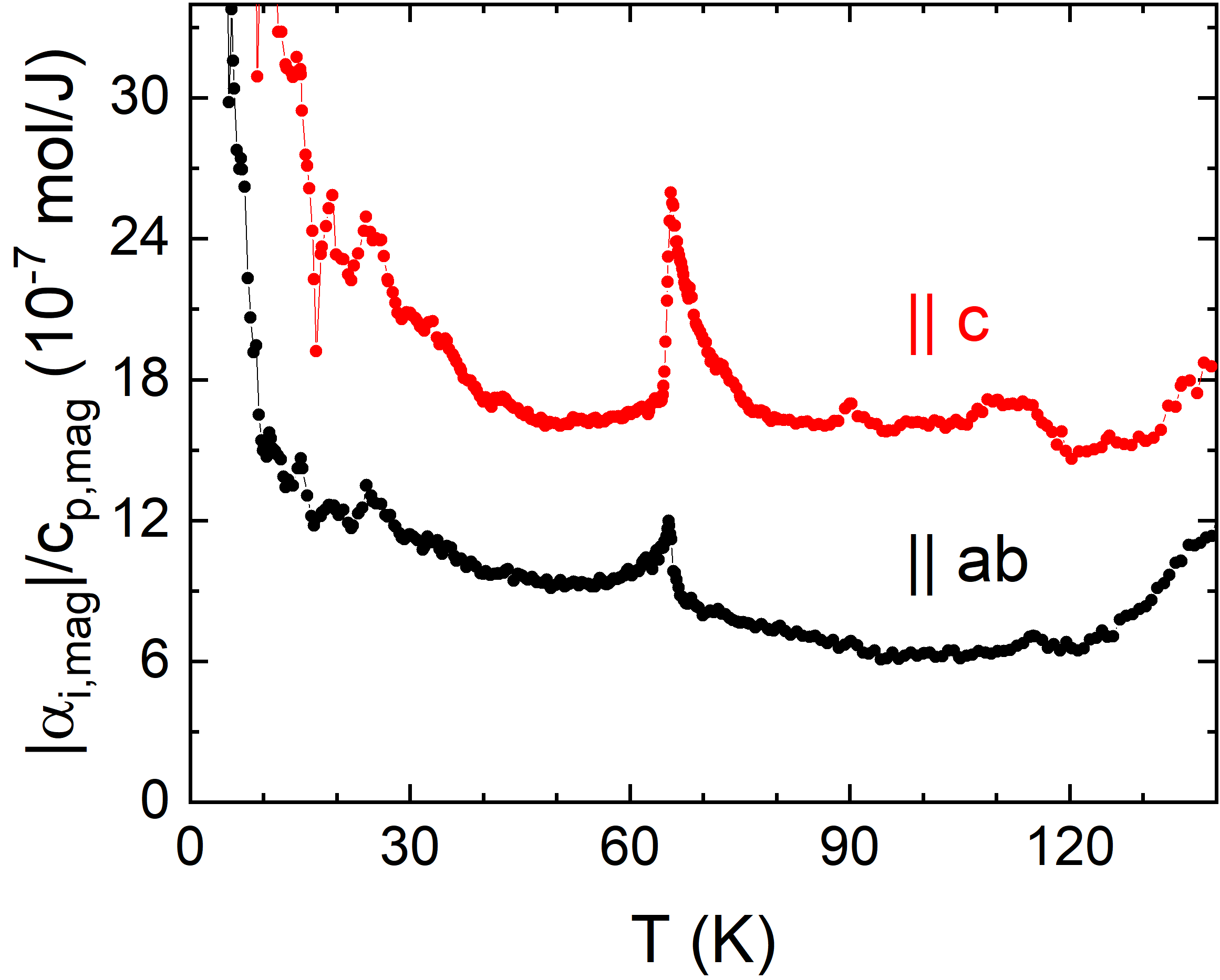}}
	\caption[] {\label{Gruen_mag} Absolute value of the effective magnetic Gr\"{u}neisen ratios $\alpha_{i,\mathrm{mag}}$/$c_{p,\mathrm{mag}}$.}
\end{figure*}

\clearpage
\section{High Temperature Magnetostriction}
Magnetostriction measurements above \TC\ $= 65$~K up to 204~K for both $B\parallel ab$ and $B\parallel c$ are shown in Fig.~\ref{SI_MS_HighT}.
It can be seen that (1) magnetostriction is of opposite sign within the $ab$ plane and along the $c$ axis, (2) magnetostriction along the $c$ axis is about 1.5 times larger than within the $ab$ plane, and (3) at around 150~K (125~K for $B \parallel c$) a sizeable magnetostriction is still present which nearly vanishes around 200~K.
The relative length changes from 0~T to 15~T from these measurements as well as measurements below \TC\ are shown in Fig.~1 in the main text.
\renewcommand{\thefigure}{S5}
\begin{figure*}[htbp]
	\center{\includegraphics [width=1\columnwidth,clip]{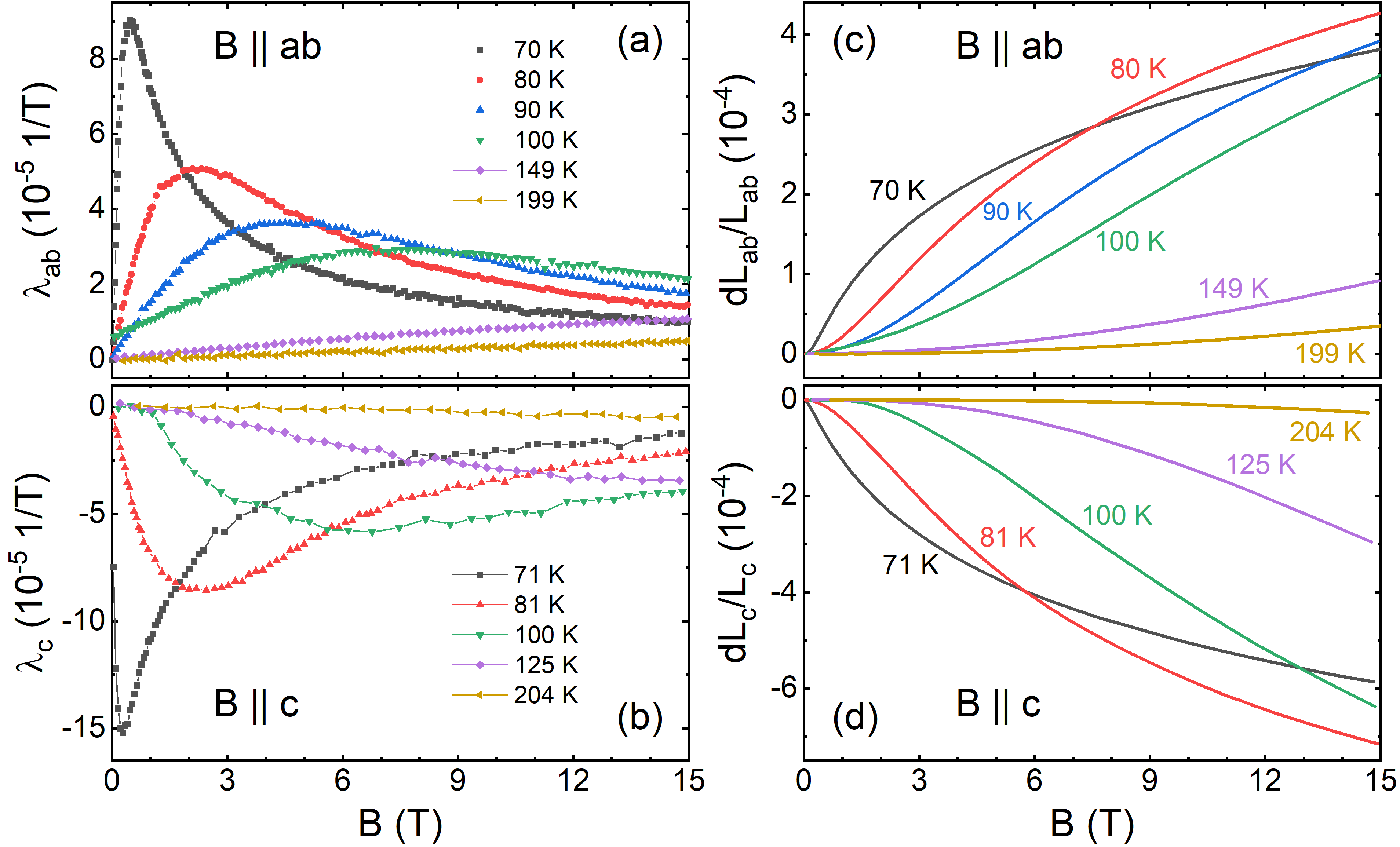}}
	\caption[] {\label{SI_MS_HighT} (a, b) Magnetostriction coefficients and (c, d) magnetostrictive relative length changes for $B\parallel ab$ (a, c) and $B\parallel c$ (b, d) at temperatures $T > T_{\mathrm{C}}$. Only up-sweep data is shown.}
\end{figure*}

\bibliography{Cr2Ge2Te6_Paper_bibliography_PRL}